\documentclass[aip,jcp,reprint]{revtex4-1}
\usepackage{amsmath,amssymb}
\usepackage{graphicx}
\usepackage{dcolumn}
\usepackage{bm}
\usepackage{multirow}
\usepackage{float}
\usepackage{subfigure}
\usepackage{color}
\usepackage{afterpage}
\usepackage{tabularx}

\usepackage[utf8]{inputenc}
\usepackage[T1]{fontenc}
\usepackage{mathptmx}
\usepackage[breaklinks]{hyperref}
\usepackage{physics}

\usepackage[dvipsnames]{xcolor}

\graphicspath{{Figures/}}

\begin{document}

\preprint{AIP/123-QED}

\title{Simple but accurate estimation of light-matter coupling strength and optical loss for a molecular emitter coupled with photonic modes}

\author{Siwei Wang}
\affiliation{Institute of Atomic and Molecular Sciences, Academia Sinica, Taipei 10617, Taiwan}
\author{Yi-Ting Chuang}
\affiliation{Institute of Atomic and Molecular Sciences, Academia Sinica, Taipei 10617, Taiwan}
\author{Liang-Yan Hsu}
\email{lyhsu@gate.sinica.edu.tw}
\affiliation{Institute of Atomic and Molecular Sciences, Academia Sinica, Taipei 10617, Taiwan}

\begin{abstract}
Light-matter coupling strength and optical loss are two key physical quantities in cavity quantum electrodynamics (cQED), and their interplay determines whether light-matter hybrid states can be formed or not in chemical systems. In this study, by using macroscopic quantum electrodynamics (mQED) combined with a pseudomode approach, we present a simple but accurate method which allows us to quickly estimate the light-matter coupling strength and optical loss without free parameters. Moreover, for a molecular emitter coupled with photonic modes (including cavity modes and plasmon polartion modes), we analytically and numerically prove that the dynamics derived from the mQED-based wavefunction approach is mathematically equivalent to the dynamics governed by the cQED-based Lindblad master equation when the Purcell factor behaves like Lorentzians. 
\end{abstract}


\maketitle

\section{Introduction}
Light-matter interaction plays a crucial role in various fields such as surface-enhanced Raman spectroscopy\cite{Fleischmann1974,Jeanmaire1977,Nitzan_1980,Weitz1983,Zhao2006,Camden2008,Morton2009,Sharma2012,Trujillo2018}, biosensing\cite{Homola2003,Hartland2011,Saha2012,Holzinger2014,Mejia-Salazar2018,Hsu2020}, photosynthesis\cite{Meyer1989,Alstrum-Acevedo2005,Scholes2011,Mirkovic2017,Jumper2018}, resonance energy transfer\cite{Andrews2004,DeTorres2016,Hsu2017,Ding2017,Hsu_2018,Rustomji2019,Lee2020}, and molecular fluorescence\cite{Drexhage1970,Chance1978,Gersten1981,Rigneault2005,Wenger2008,Aouani2011,Wang2019,Wang2020,Wang2020a}. Recently, due to advances in nanotechnology, the achievement of strong light-matter interaction\cite{Vasa2013,Chikkaraddy2016,Han2018,Beane2018,Yadav2020} has motivated successive studies on the chemical properties of light–matter hybrid states, leading to the emergence of polariton chemistry\cite{Schwartz2011,Ebbesen2016,Galego2017,Campos-Gonzalez-Angulo2019,Thomas2019}. To investigate how strong light-matter interactions influence chemical reactions, most studies construct theoretical models based on cavity quantum electrodynamics (cQED), including Jaynes-Cummings model\cite{Ribeiro2018,Yuen-Zhou2019}, Holstein–Tavis–Cummings model\cite{Spano2015,Spano2020}, and other related model Hamiltonians\cite{Vendrell2018,Semenov2019}. However, the light-matter coupling strengths in these models are obtained by fitting experimental data instead of theoretical calculations because of the difficulty in estimating the mode volume\cite{craig1998molecular,scully1999quantum}. Apart from the coupling strength, optical loss also plays an important role in the formation of the light-matter hybrid states, especially in dispersive and absorbing environments\cite{Herrera2018,Antoniou2020}, e.g., metallic nanocavities. Nevertheless, it is a challenge to estimate optical loss in theory because this quantity is related to local dielectric environments. Instead, most theoretical studies regard optical loss as a free parameter\cite{Yuge2014,Davidsson2020} or obtain them from experiments \cite{Herrera2017,Herrera2017_2,Ulusoy2020}. The lack of rigorous consideration of the two physical quantities may lead to experimental misunderstandings and hinder us from quantitatively predicting polariton-coupled chemical and physical processes. Therefore, it is paramount to establish a rigorous theory which enables us to quantitatively estimate light-matter coupling strength and optical loss in complicated dielectric environments\cite{Hu2006}.

To incorporate the effect of dielectric environments, we developed a general theory of a molecular emitter strongly coupled with plasmon polaritons in the framework of macroscopic quantum electrodynamics (mQED)\cite{Wang2019,Wang2020}. In addition, in the high-frequency limit, we derived a parameter-free formula which can be used to estimate light-molecule coupling strengths, and the coupling strength given by this parameter-free formula is in agreement with the experimental results done by Cavendish Laboratory\cite{Chikkaraddy2016}. However, the previous work does not rigorously prove how mQED turns into the form of cQED, nor does it emphasize how to calculate optical loss from mQED. In this article, we address these two issues by combining mQED and a pseudomode approach, which replaces a reservoir of infinite photonic modes with a few discrete Lindblad-damped harmonic oscillators\cite{Imamoglu1994,Garraway1996,Garraway1997,Dalton2001,Gonzalez-Tudela2014,Pleasance2020,Mascherpa2020}. In other words, we establish a theory which bridges our previous mQED-based wavefunction approach and a cQED-based Lindblad master equation. Here, we would like to emphasize that the cQED-based Lindblad master equation can be used to describe a molecular emitter strongly coupled with photonic modes because the pseudomode approach is valid from the weak to strong light-matter coupling regimes\cite{Garraway1997,Gonzalez-Tudela2014}. Moreover, through the present theory, one can simply and accurately estimate light-matter coupling strength and optical loss for a molecular emitter with vibrational degrees of freedom in complicated dielectric environments.

Our article is organized as follows. In Sec.~\ref{Sec:Method}, we briefly recapitulate our previous study about the quantum dynamics of molecular fluorescence based on mQED. According to our previous study,
we establish a rigorous theory which allows us to quantitatively estimate the two key physical quantities. 
In Sec.~\ref{Sec:Result}, in order to clearly show the use and advantages of our theory, we investigate two representative systems: a spherical silver cavity (cavity photon) and a NaCl-coated silver surface (plasmon polariton). In Sec.~\ref{Sec:Conclusion}, we summarize the main
results and provide a perspective for future
work.

\section{Theory}
\label{Sec:Method}

\subsection{Quantum dynamics based on mQED theory}
\label{Sec:Hamiltonian}
The Hamiltonian of a two-electronic-state molecule with multi-vibrational modes coupled to a vacuum field in a dielectric environment can be expressed as\cite{Wang2020,Wang2020a}
\begin{align}
    \hat{H}\equiv & \ket{\mathrm{g}}\left( \sum_{u=1}^{n_\mathrm{vib}} \hbar\omega_{\mathrm{vib},u}\left[ \hat{b}_u^\dag \hat{b}_u -\sqrt{S_u}(\hat{b}_u^\dag+\hat{b}_u)+S_u  \right]  \right)\bra{\mathrm{g}} \nonumber \\
    &+ \ket{\mathrm{e}}\left( \hbar\omega_{\mathrm{eg}} + \sum_{u=1}^{n_\mathrm{vib}} \hbar\omega_{\mathrm{vib},u} \hat{b}_u^\dag \hat{b}_u \right)\bra{ \mathrm{e}} \nonumber \\
    &+\int d\mathbf{r}\int_0^\infty d\omega \, \hbar\omega \, \hat{\mathbf{f}}^\dag(\mathbf{r},\omega)\cdot\hat{\mathbf{f}}(\mathbf{r},\omega) \nonumber \\
    &-\left( \ket{\mathrm{e}}\bra{\mathrm{g}}\hat{\mathbf{E}}^{(+)}(\mathbf{r}_\mathrm{M})\cdot\boldsymbol{\mu} + \ket{\mathrm{g}}\bra{\mathrm{e}}\hat{\mathbf{E}}^{(-)}(\mathbf{r}_\mathrm{M})\cdot\boldsymbol{\mu}  \right),
\label{Eq:Working_Hamiltonian}
\end{align}
where $\mathbf{r}_\mathrm{M}$ is the position of the molecule in the dielectric environment. $\ket{\mathrm{g}}$ ($\ket{\mathrm{e}}$) is the electronically ground (excited) state of the molecule. The energy gap $\hbar\omega_\mathrm{eg}$ corresponds to the adiabatic excitation energy from the electronically ground state to the electronically excited state, and $\boldsymbol{\mu}$ stands for the transition dipole moment of the molecule. The symbol $u$ denotes the index of the harmonic vibrational modes from $1$ to $n_\mathrm{vib}$. $\hat{b}_u^\dagger$ ($\hat{b}_u$), $\omega_{\mathrm{vib},u}$, and $S_\mathrm{u}$ are the creation (annihilation) operator, molecular vibrational frequency, and Huang-Rhys factor of the $u$-th vibrational mode, respectively. $\hat{\mathbf{E}}^{(+)}(\mathbf{r}_\mathrm{M})$ is the summation of the positive-frequency electric field operators and is related to the annihilation operator for bosonic vector fields (polariton) $\hat{\mathbf{f}}(\mathbf{r},\omega)$ according to the mQED theory\cite{Gruner1996,Dung2000,Ritter2018,Hemmerich2018,Lindel2021}, i.e., $\hat{\mathbf{E}}^{(+)}(\mathbf{r}_\mathrm{M}) \equiv i\int_0^\infty d\omega \sqrt{\frac{\hbar}{\pi\epsilon_0}}  \frac{\omega^2}{c^2}\int d\mathbf{r}\sqrt{\mathrm{Im}\left\{\epsilon_\mathrm{r}(\mathbf{r},\omega)\right\}} \overline{\overline{\mathbf{G}}}(\mathbf{r}_\mathrm{M},\mathbf{r},\omega)\cdot\hat{\mathbf{f}}(\mathbf{r},\omega)$, where $\epsilon_0$ is the vacuum permittivity, $c$ is the speed of light in vacuum, and $\mathrm{Im}\left\{\epsilon_\mathrm{r}(\mathbf{r},\omega)\right\}$ is the imaginary part of the complex dielectric function of the environment. In addition, $\hat{\mathbf{E}}^{(-)}(\mathbf{r}_\mathrm{M})$ is the Hermitian conjugate of $\hat{\mathbf{E}}^{(+)}(\mathbf{r}_\mathrm{M})$. The dyadic Green's function $\overline{\overline{\mathbf{G}}}(\mathbf{r}_\mathrm{M},\mathbf{r},\omega)$ satisfies Maxwell's equations $\left((\omega/c)^2\epsilon_\mathrm{r}(\mathbf{r}_\mathrm{M},\omega) - \nabla \times \nabla \times \right)\overline{\overline{\mathbf{G}}}(\mathbf{r}_\mathrm{M},\mathbf{r},\omega)=-\mathbf{\overline{\overline{I}}}_3\delta(\mathbf{r}_\mathrm{M}-\mathbf{r})$\cite{chew1995waves,Novotny2012}, where $\mathbf{\overline{\overline{I}}}_3$ is a $3\times3$ identity matrix and $\delta(\mathbf{r}_\mathrm{M}-\mathbf{r})$ is the three-dimensional delta function.

To study quantum dynamics of molecular fluorescence (spontaneous emission), we use a wavefunction ansatz based on the
Wigner-Weisskopf theory as follows\cite{Wang2020,Wang2020a},
{\small  
\begin{align}
\nonumber
\ket{\psi(t)}=&   \sum_{\left\{M\right\}_\mathrm{vib}=0}^\infty  C^\mathrm{e,\left\{0\right\}}_{\left\{M\right\}_\mathrm{vib}}(t)e^{-i(\omega_\mathrm{eg}+ \sum_{u=1}^{n_\mathrm{vib}} \omega_{\mathrm{vib},u}M_u )t}\ket{\mathrm{e}}\ket{\{0\}} \prod_{u=1}^{n_\mathrm{vib}}\ket{M_u(0)}\\
\nonumber
&+ \sum_{i=1}^3\sum_{\left\{M'\right\}_\mathrm{vib}=0}^\infty \int d\mathbf{r}\int_0^\infty d\omega\, C^{\mathrm{g},\left\{1_i\right\}}_{\left\{M'\right\}_\mathrm{vib}}(\mathbf{r},\omega,t)e^{-i(\omega+\sum_{u=1}^{n_\mathrm{vib}}\omega_{\mathrm{vib},u}M'_u)t}\\
&\times \ket{\mathrm{g}} \ket{\left\{1_{i}(\mathbf{r},\omega)\right\}} \prod_{u=1}^{n_\mathrm{vib}}\ket{M'_u(S_u)}, 
\label{Wavefunction_Ansatz}
\end{align}
}where $\ket{\left\{0\right\}}$ is the vacuum state of the electromagnetic field, and $\ket{\left\{1_{i}(\mathbf{r},\omega)\right\}}$ denotes a Fock state of one $i$-directional polarized polariton with a certain frequency $\omega$ at position $\mathbf{r}$. Note that $\ket{\left\{1_{i}(\mathbf{r},\omega)\right\}}$ and $\ket{\left\{0\right\}}$ are associated via the bosonic vector fields $\hat{\mathbf{f}}^\dagger(\mathbf{r},\omega)=\left(\hat{\mathrm{f}}_x^\dagger(\mathbf{r},\omega),\hat{\mathrm{f}}_y^\dagger(\mathbf{r},\omega),\hat{\mathrm{f}}_z^\dagger(\mathbf{r},\omega)\right)$, i.e., $\ket{\left\{1_{i}(\mathbf{r},\omega)\right\}}=\hat{\mathrm{f}}_i^\dagger(\mathbf{r},\omega)\ket{\left\{0\right\}}$. $\sum_{i=1}^3$ corresponds to the summation of all polarization directions $x,\,y,\,z$. $\ket{M_u(0)}$ is the eigenstate of the $u$-th normal mode in the molecular excited state, i.e., $\hat{b}_u^\dag \hat{b}_u \ket{M_u(0)} = M_u \ket{M_u(0)}$. Similarly, the eigenstate of the $u$-th normal mode in the molecular ground state is the displaced Fock state $\ket{M'_u(S_u)}$, i.e., $\left[ \hat{b}_u^\dag \hat{b}_u -\sqrt{S_u}(\hat{b}_u^\dag+\hat{b}_u)+S_u  \right] \ket{M'_u(S_u)} = M'_u\ket{M'_u(S_u)}$. $C^\mathrm{e,\left\{0\right\}}_{\left\{M\right\}_\mathrm{vib}}(t)$ is the coefficient of the molecular excited state (the superscript $\mathrm{e}$) with the vacuum state of electromagnetic field (the superscript $\left\{0\right\}$) having the quanta of $n_\mathrm{vib}$ different vibrational modes (the subscripts $\left\{M\right\}_\mathrm{vib}=M_1, M_2,\dots,M_{n_\mathrm{vib}}$). $\sum_{\left\{M\right\}_\mathrm{vib}=0}^\infty$ means the summation of all vibrational modes and all vibrational quanta. Similarly, $C^{\mathrm{g},\left\{1_i\right\}}_{\left\{M'\right\}_\mathrm{vib}}(t)$ is the coefficient of molecular ground state (the superscript $\mathrm{g}$) with a polariton having a specific $i$-polarization direction.

According to our previous works \cite{Wang2020,Wang2020a}, we have proven that the Schr\"{o}dinger equation with the Hamiltonian in Eq.~(\ref{Eq:Working_Hamiltonian}) and $\ket{\psi(t)}$ in Eq.~(\ref{Wavefunction_Ansatz}) can turn into the following integro-differential equations in terms of the coefficient $C^\mathrm{e,\left\{0\right\}}_{\left\{M\right\}_\mathrm{vib}}(t)$,
{\small
\begin{align}
    \frac{d C_{\left\{K\right\}_\mathrm{vib}}^\mathrm{e,\left\{0\right\}}(t)}{dt}=&-\sum_{\left\{M\right\}_\mathrm{vib}=0}^\infty \int_0^t dt' K_\mathrm{pol}(t,t')    K_\mathrm{vib}^{ \left\{K\right\}_\mathrm{vib}\leftarrow\left\{M\right\}_\mathrm{vib}}(t,t') C_{\left\{M\right\}_\mathrm{vib}}^\mathrm{e,\left\{0\right\}}(t'),
    \label{Eq:Dynamic_Equation}
\end{align}
}where $K_\mathrm{pol}(t,t')$ is the memory kernel of polariton which has been defined in our previous studies\cite{Wang2019,Wang2020,Wang2020a},
\begin{align}
    K_\mathrm{pol}(t,t')
    & = \frac{1}{2\pi} \int_0^\infty d\omega\; A_0(\omega)F_\mathrm{P}(\omega) e^{-i(\omega-\omega_\mathrm{eg})(t-t')},
\label{Eq:Polariton_Kernel}
\end{align}
where $F_\mathrm{P}(\omega)$ is the Purcell factor at a certain frequency $\omega$. The Purcell factor can be expressed as the ratio of $A_\mathrm{FG}(\omega)$ to $A_0(\omega)$, i.e., $F_\mathrm{P}(\omega)={A_\mathrm{FG}(\omega)}/{A_0(\omega)}$.  These two emission rates are well-known as\cite{Novotny2012},
\begin{align}
    &A_\mathrm{FG}(\omega) = \frac{2\omega^2}{\hbar c^2\epsilon_0}  \boldsymbol{\mu} \cdot  \mathrm{Im}\left\{\overline{\overline{\mathbf{G}}}(\mathbf{r}_\mathrm{M},\mathbf{r}_\mathrm{M},\omega) \right\} \cdot \boldsymbol{\mu},
    \label{Eq:Media_Decay_Rate}
    \\    
    &A_0(\omega) = \frac{\omega^3\abs{\boldsymbol{\mu}}^2}{3c^3\hbar\pi\epsilon_0},
    \label{Eq:Vacuum_Decay_Rate}
\end{align}
where $A_\mathrm{FG}(\omega)$ corresponds to the spontaneous emission rate in a dielectric environment, and $A_0(\omega)$ represents the spontaneous emission rate in vacuum.

Based on our previous works\cite{Wang2020,Wang2020a}, the memory kernel of molecular vibrations $K_\mathrm{vib}^{ \left\{K\right\}_\mathrm{vib}\leftarrow\left\{M\right\}_\mathrm{vib}}(t,t')$ can be expressed as 
\begin{align}
\nonumber
    K_\mathrm{vib}^{ \left\{K\right\}_\mathrm{vib}\leftarrow\left\{M\right\}_\mathrm{vib}}(t,t')  =&\prod_{u=1}^{n_\mathrm{vib}} \sum_{M'_u=0}^\infty \braket{K_u(0)}{M'_u(S_u)}e^{i\omega_{\mathrm{vib},u}(K_u-M'_u)t}\\
    & \braket{M'_u(S_u)}{M_u(0)}e^{i\omega_{\mathrm{vib},u}(M'_u-M_u)t'}, 
    \label{Eq:Vib-Kernel}
\end{align}
where $\braket{K_u(0)}{M'_u(S_u)}$ is the vibrational overlap of the electronically excited state and the electronically ground state. For independent quantum harmonic oscillators, the vibrational overlap depends on the Huang-Rhys factor $S_u$\cite{Wang2020a}. To summarize, Eqs.~(\ref{Eq:Dynamic_Equation}), (\ref{Eq:Polariton_Kernel}), and (\ref{Eq:Vib-Kernel}) are the working equations for the quantum dynamics of a molecule coupled with dielectric environments.

\subsection{Pseudomode approach and the corresponding Lindblad master equation}

The mapping of the quantum dynamics from the mQED theory to the cQED model relies on two premises: one is the validity of the flat continuum approximation\cite{Thanopulos2017,Wang2019}, and the other is the Lorentz-shape Purcell factor, 
\begin{align}
    F_\mathrm{P}(\omega) = \sum_{j=1}^{n_\mathrm{p}} \frac{F_j\,\Gamma_j^2}{(\omega-\omega_j)^2+\Gamma_j^2},
    \label{Eq:Lorentz}
\end{align}
where $n_\mathrm{p}$ is the number of Lorentzian functions used to fit the Purcell factor. $\omega_j$, $F_j$, and $\Gamma_j$ correspond to the peak frequency, the peak height, and the half width at half maximum of the $j$-th Lorentzian function, respectively.

Substituting Eq.~(\ref{Eq:Lorentz}) into the memory kernel of polariton Eq.~(\ref{Eq:Polariton_Kernel}), we can obtain
\begin{widetext}
\begin{align}
\nonumber
    K_\mathrm{pol}(t,t')
    &= \sum_{j=1}^{n_\mathrm{p}} \int_0^\infty d\omega\; \frac{A_0(\omega)}{2\pi} \frac{F_j\,\Gamma_j^2}{(\omega-\omega_j)^2+\Gamma_j^2} e^{-i(\omega-\omega_\mathrm{eg})(t-t')} \\
    &  \approx \sum_{j=1}^{n_\mathrm{p}}  \frac{A_0(\omega_j)}{2\pi} \int_0^\infty d\omega\;  \frac{F_j\,\Gamma_j^2}{(\omega-\omega_j)^2+\Gamma_j^2} e^{-i(\omega-\omega_\mathrm{eg})(t-t')} \label{Eq:flat-continum_app} \\
    &  \approx \sum_{j=1}^{n_\mathrm{p}}  \frac{A_0(\omega_j)}{2\pi} \int_{-\infty}^\infty d\omega\;  \frac{F_j\,\Gamma_j^2e^{-i(\omega-\omega_\mathrm{eg})(t-t')} }{(\omega-\omega_j+i\Gamma_j)(\omega-\omega_j-i\Gamma_j)} \label{Eq:Extend_lower} \\
    & =\sum_{j=1}^{n_\mathrm{p}} \frac{A_0(\omega_j)F_j\Gamma_j}{2}  e^{-i(\omega_j-\omega_\mathrm{eg}-i\Gamma_j)(t-t')}.
    \label{Eq:Kernel_Result}
\end{align}

Note that we use the flat continuum approximation $A_0(\omega)\approx A_0(\omega_j)$ to derive Eq.~(\ref{Eq:flat-continum_app}) because $A_0(\omega)$ in Eq.~(\ref{Eq:Vacuum_Decay_Rate}) can be regarded as slowly varying near $\omega_j$ compared with $F_\mathrm{P}(\omega)$ in Eq.~(\ref{Eq:Lorentz}) when $\Gamma_j/\omega_j\ll 1$. In Eq.~(\ref{Eq:Extend_lower}), we extend the lower limit to $-\infty$ due to the fact that $F_\mathrm{P}(\omega)\rightarrow 0$ when $\omega<0$. Therefore, we can obtain Eq.~(\ref{Eq:Kernel_Result}) by using a contour integration of Eq.~(\ref{Eq:Extend_lower}), where the path of the contour is taken to be a semicircle in the lower half-plane. Substituting Eq.~(\ref{Eq:Kernel_Result}) into the dynamical equation Eq.~(\ref{Eq:Dynamic_Equation}), we obtain
\begin{align}
    \frac{d C_{\left\{K\right\}_\mathrm{vib}}^\mathrm{e,\left\{0\right\}}(t)}{dt}=&-\sum_{j=1}^{n_\mathrm{p}} \sum_{\left\{M\right\}_\mathrm{vib}=0}^\infty \int_0^t dt' \left\{ \frac{A_0(\omega_j)F_j\Gamma_j}{2}  e^{-i(\omega_j-\omega_\mathrm{eg}-i\Gamma_j)(t-t')} \right\} 
    \nonumber \\ &\times
    \left\{ \prod_{u=1}^{n_\mathrm{vib}} \sum_{M'_u=0}^\infty \braket{K_u(0)}{M'_u(S_u)} e^{i\omega_{\mathrm{vib},u}(K_u-M'_u)t} \braket{M'_u(S_u)}{M_u(0)}e^{i\omega_{\mathrm{vib},u}(M'_u-M_u)t'} \right\} C_{\left\{M\right\}_\mathrm{vib}}^\mathrm{e,\left\{0\right\}}(t').
\label{Eq:12}
\end{align}

According to the concept of the pseudomode method\cite{Imamoglu1994,Garraway1996,Garraway1997,Dalton2001,Gonzalez-Tudela2014,Pleasance2020,Mascherpa2020}, we can define the pseudomode amplitude as
\begin{align}
    B^{\mathrm{g},\left\{1_j\right\}_\mathrm{p}}_{\left\{M'\right\}_\mathrm{vib}} (t) = i \sum_{\left\{M\right\}_\mathrm{vib}=0}^\infty \int_0^t dt' \sqrt{\frac{A_0(\omega_j)F_j\Gamma_j}{2}}  C_{\left\{M\right\}_\mathrm{vib}}^\mathrm{e,\left\{0\right\}}(t') e^{-i(\omega_\mathrm{eg}-\omega_j+i\Gamma_j+\sum_{u=1}^{n_\mathrm{vib}}(M_u-M'_u))t'} \prod_{u=1}^{n_\mathrm{vib}}\braket{M'_u(S_u)}{M_u(0)}.
\label{Eq:13}
\end{align}

Substituting Eq.~(\ref{Eq:13}) into Eq.~(\ref{Eq:12}),  Eq.~(\ref{Eq:12}) can be expressed as
\begin{align}
    \frac{d C_{\left\{K\right\}_\mathrm{vib}}^\mathrm{e,\left\{0\right\}}(t)}{dt}=&  i\sum_{j=1}^{n_\mathrm{p}}\sum_{\left\{M'\right\}_\mathrm{vib}=0}^\infty  \left\{ \sqrt{ \frac{A_0(\omega_j)F_j\Gamma_j}{2} }  e^{-i(\omega_j-\omega_\mathrm{eg}-i\Gamma_j)t} \right\}  \left\{ \prod_{u=1}^{n_\mathrm{vib}}  \braket{K_u(0)}{M'_u(S_u)}e^{i\omega_{\mathrm{vib},u}(K_u-M'_u)t}  \right\} B_{\left\{M'\right\}_\mathrm{vib}}^\mathrm{g,\left\{1_j\right\}_\mathrm{p}}(t).
\label{Eq:14}
\end{align}

Similarly, according to Eq.~(\ref{Eq:13}), the differential equation of the pseudomode amplitude $B^{\mathrm{g},\left\{1\right\}}_{\left\{M'\right\}_\mathrm{vib}} (t)$ can be written as
\begin{align}
    \frac{dB^{\mathrm{g},\left\{1_j\right\}_\mathrm{p}}_{\left\{M'\right\}_\mathrm{vib}} (t)}{dt} = i \sum_{\left\{M\right\}_\mathrm{vib}=0}^\infty \left\{ \sqrt{\frac{A_0(\omega_j)F_j\Gamma_j}{2}}   e^{-i(\omega_\mathrm{eg}-\omega_j+i\Gamma_j)t} \right\} \left\{ \prod_{u=1}^{n_\mathrm{vib}}\braket{M'_u(S_u)}{M_u(0)} e^{i\omega_{\mathrm{vib},u}(M'_u-M_u)t} \right\} C_{\left\{M\right\}_\mathrm{vib}}^\mathrm{e,\left\{0\right\}}(t).
\label{Eq:15}
\end{align}

Note that $C_{\left\{K\right\}_\mathrm{vib}}^\mathrm{e,\left\{0\right\}}(t)$ and $B^{\mathrm{g},\left\{1_j\right\}_\mathrm{p}}_{\left\{M'\right\}_\mathrm{vib}} (t)$ can be regarded as a projection from the total wavefunction to the pseudomode basis $\ket{\mathrm{e},\left\{0\right\}_\mathrm{p}}\prod_{u=1}^{n_\mathrm{vib}}\ket{K_u(0)}$ and $\ket{\mathrm{g},\left\{1_j\right\}_\mathrm{p}}\prod_{u=1}^{n_\mathrm{vib}}\ket{M'_u(S_u)}$, respectively, where $\left\{0\right\}_\mathrm{p}$ denotes 0 quanta for all pseudomodes and $\left\{1_j\right\}_\mathrm{p}$ denotes only the $j$-th pseudomode with $1$ quantum (the others with 0 quanta) in the set of pseudomodes. Inspired by the projection and the form of Eqs.~(\ref{Eq:14}) and (\ref{Eq:15}), we can write down the total wavefunction based on the wavefunction ansatz in terms of the pseudomode basis,
{\small  
\begin{align}
\label{Eq:16}
    &\ket{\varPhi(t)} = \sum_{\left\{M\right\}_\mathrm{vib}=0}^\infty C_{\left\{M\right\}_\mathrm{vib}}^\mathrm{e,\left\{0\right\}}(t) e^{-i(\omega_\mathrm{eg}+\sum_{u=1}^{n_\mathrm{vib}}\omega_{\mathrm{vib},u}M_u)t} \ket{\mathrm{e},\left\{0\right\}_\mathrm{p}} \prod_{u=1}^{n_\mathrm{vib}}\ket{M_u(0)}+\sum_{j=1}^{n_\mathrm{p}}\sum_{\left\{M'\right\}_\mathrm{vib}=0}^\infty B_{\left\{M'\right\}_\mathrm{vib}}^{\mathrm{g},\left\{1_j\right\}_\mathrm{p}}(t) e^{-i(\omega_j-i\Gamma_j+\sum_{u=1}^{n_\mathrm{vib}}\omega_{\mathrm{vib},u}M'_u)t} \ket{\mathrm{g},\left\{1_j\right\}_\mathrm{p}} \prod_{u=1}^{n_\mathrm{vib}}\ket{M'_u(S_u)}.
\end{align}
} 
\end{widetext}

By using this wavefunction ansatz $\ket{\varPhi(t)}$, we can construct an effective non-Hermitian Hamiltonian $\hat{H}_\mathrm{eff}$, which replicates the same equations as Eq.~(\ref{Eq:14}) and Eq.~(\ref{Eq:15}) with the Schr\"{o}dinger equation $ i\hbar \frac{d}{dt} \ket{\varPhi(t)} =\hat{H}_\mathrm{eff} \ket{\varPhi(t)}$ (the details can be found in Appendix~\ref{Appendix_A}),
\begin{align}
\nonumber
    \hat{H}_\mathrm{eff} \equiv  
    & \ket{\mathrm{g}}\left( \sum_{u=1}^{n_\mathrm{vib}}  \hbar\omega_{\mathrm{vib},u}\left[ b_u^\dag b_u -\sqrt{S_u}(b_u^\dag+b_u)+S_u \right]  \right)\bra{\mathrm{g}}\\
    &+ \ket{\mathrm{e}}\left( \hbar\omega_{\mathrm{eg}} + \sum_{u=1}^{n_\mathrm{vib}}  \hbar\omega_{\mathrm{vib},u} b_u^\dag b_u \right)\bra{ \mathrm{e}} \nonumber \\
    &- \sum_{j=1}^{n_\mathrm{p}} V_j\left( \ket{\mathrm{e}}\bra{\mathrm{g}}\hat{a}_j + \ket{\mathrm{g}}\bra{\mathrm{e}}\hat{a}^+_j   \right) + \sum_{j=1}^{n_\mathrm{p}} \hbar(\omega_j-i\Gamma_j) \hat{a}_j^+\hat{a}_j,
\label{Eq:Effective_Hamiltonian}
\end{align}
where $\hat{a}_j^\dagger$ and $\hat{a}_j$ are the creation and annihilation operator of the $j$-th pseudomode, i.e., $\hat{a}_j\ket{\left\{1_j\right\}_\mathrm{p}}=\ket{\left\{0\right\}_\mathrm{p}}$ and $\hat{a}^\dagger_j\ket{\left\{0\right\}_\mathrm{p}}=\ket{\left\{1_j\right\}_\mathrm{p}}$, and $V_j=\hbar\sqrt{{A_0(\omega_j)F_j\Gamma_j}/{2}}$ is the coupling strength between the molecule and the $j$-th pseudomode. 

In order to transform this quantum dynamics into a standard form of Lindblad master equation, we first define an effective density matrix, which is composed of the wavefunction ansatz, i.e., $\hat{\rho}_\mathrm{eff}(t) = \ket{\varPhi(t)}\bra{\varPhi(t)}$. Obviously, it satisfies the following von Neumann equation,
\begin{align}
    i\hbar\frac{d}{dt}\hat{\rho}_\mathrm{eff}(t) = \hat{H}_\mathrm{eff}\hat{\rho}_\mathrm{eff}(t) - \hat{\rho}_\mathrm{eff}(t)\hat{H}_\mathrm{eff}^\dagger \equiv \left[\hat{H}_\mathrm{eff},\hat{\rho}_\mathrm{eff}(t)\right].
\label{Eq:20}
\end{align}

Due to the non-Hermiticity of $\hat{H}_\mathrm{eff}$ in Eq.~(\ref{Eq:20}), we separate $\hat{H}_\mathrm{eff}$ into a Hermitian part $\hat{H}_0$ and a non-Hermitian part $-i\hbar\sum_{j=1}^{n_\mathrm{p}}\Gamma_j\hat{a}_j^\dagger\hat{a}_j$, i.e., $\hat{H}_\mathrm{eff} = \hat{H}_0 - i\hbar\sum_{j=1}^{n_\mathrm{p}}\Gamma_j\hat{a}_j^\dagger\hat{a}_j$. Moreover, $\hat{H}_0$ can be simplified to the form of a cQED Hamiltonian,
\begin{align}
\nonumber
    \hat{H}_\mathrm{0} \equiv  
    & \ket{\mathrm{g}}\left( \sum_{u=1}^{n_\mathrm{vib}} \hbar\omega_{\mathrm{vib},u}\left[ b_u^\dag b_u -\sqrt{S_u}(b_u^\dag+b_u)+S_u \right]  \right)\bra{\mathrm{g}}\\
    &+ \ket{\mathrm{e}}\left( \hbar\omega_{\mathrm{eg}} + \sum_{u=1}^{n_\mathrm{vib}} \hbar\omega_{\mathrm{vib},u} b_u^\dag b_u \right)\bra{ \mathrm{e}} \nonumber \\
    &- \sum_{j=1}^{n_\mathrm{p}} {V_j}\left( \ket{\mathrm{e}}\bra{\mathrm{g}}\hat{a}_j + \ket{\mathrm{g}}\bra{\mathrm{e}}\hat{a}^+_j   \right) + \sum_{j=1}^{n_\mathrm{p}} \hbar\omega_j \hat{a}_j^+\hat{a}_j,
\label{Eq:H0}
\end{align}
and Eq.~(\ref{Eq:20}) can be rearranged as follows 
\begin{align}
\nonumber
    \frac{d}{dt}\hat{\rho}_\mathrm{eff}(t) &=\frac{-i}{\hbar} \left[\hat{H}_\mathrm{0}-i\hbar\sum_{j=1}^{n_\mathrm{p}}\Gamma_j\hat{a}_j^\dagger\hat{a}_j,\hat{\rho}_\mathrm{eff}(t)\right] \\
    &= \frac{-i}{\hbar} \left[\hat{H}_\mathrm{0},\hat{\rho}_\mathrm{eff}(t)\right] - \sum_{j=1}^{n_\mathrm{p}}\Gamma_j \left( \hat{a}_j^\dagger\hat{a}_j\hat{\rho}_\mathrm{eff}(t) + \hat{\rho}_\mathrm{eff}(t)\hat{a}_j^\dagger\hat{a}_j \right).
\label{Eq:76}
\end{align}

Furthermore, in order to convert Eq.~(\ref{Eq:76}) into the standard form of the Lindblad master equation, we can add an additional term $\sum_{j=1}^{n_\mathrm{p}}2\Gamma\hat{a}_j\hat{\rho}_\mathrm{eff}(t)\hat{a}_j^\dagger$ into Eq.~(\ref{Eq:76}). Note that this additional term does not affect the quantum dynamics of $\hat{\rho}_\mathrm{eff}(t)$ because it is only related to the molecular ground state together with the vacuum state of the pseudomodes, i.e.,
{\small
\begin{align}
\nonumber
   &\sum_{j=1}^{n_\mathrm{p}}2\Gamma\hat{a}_j\hat{\rho}_\mathrm{eff}(t)\hat{a}_j^\dagger = \sum_{j=1}^{n_\mathrm{p}} \left\{  \sum_{\left\{M'\right\}_\mathrm{vib}=0}^\infty \sum_{\left\{M''\right\}_\mathrm{vib}=0}^\infty  B_{\left\{M'\right\}_\mathrm{vib}}^{\mathrm{g},\left\{1_j\right\}_\mathrm{p}}(t) B_{\left\{M''\right\}_\mathrm{vib}}^{\mathrm{g},\left\{1_j\right\}_\mathrm{p}*}(t) \right\} \\
    &\times \ket{\mathrm{g},\left\{0\right\}_\mathrm{p}}\bra{\mathrm{g},\left\{0\right\}_\mathrm{p}}  \otimes \left\{ 2\Gamma e^{-i\sum_{u=1}^{n_\mathrm{vib}}\omega_{\mathrm{vib},u}(M'_u-M''_u)t}   \prod_{u=1}^{n_\mathrm{vib}}\ket{M'_u(S_u)}\bra{M''_u(S_u)} \right\} \nonumber \\
    &\propto \ket{\mathrm{g},\left\{0\right\}_\mathrm{p}}\bra{\mathrm{g},\left\{0\right\}_\mathrm{p}}.
\label{Eq:additional_term}
\end{align}} 

Recall that $\hat{\rho}_\mathrm{eff}(t)$ is associated with the single excitation of electronic or photonic manifolds (i.e., $\hat{\rho}_\mathrm{eff}(t) \propto \sum_{j=1}^{n_\mathrm{p}} \ket{\mathrm{e},\left\{0\right\}_\mathrm{p}}\bra{\mathrm{g},\left\{1_j\right\}_\mathrm{p}} +\ket{\mathrm{g},\left\{1_j\right\}_\mathrm{p}}\bra{\mathrm{g},\left\{1_j\right\}_\mathrm{p}}  + \ket{\mathrm{g},\left\{1_j\right\}_\mathrm{p}}\bra{\mathrm{e},\left\{0\right\}_\mathrm{p}} +\ket{\mathrm{e},\left\{0\right\}_\mathrm{p}}\bra{\mathrm{e},\left\{0\right\}_\mathrm{p}} $). As a result, $\sum_{j=1}^{n_\mathrm{p}}2\Gamma\hat{a}_j\hat{\rho}_\mathrm{eff}(t)\hat{a}_j^\dagger$ is decoupled 
from the dynamics of $\hat{\rho}_\mathrm{eff}(t)$.  After adding this additional term into Eq.~(\ref{Eq:76}), the equation of motion of $\hat{\rho}_\mathrm{eff}(t)$ can be cast into a standard Lindblad form,
\begin{align}
\nonumber
    \frac{d}{dt}\hat{\rho}_\mathrm{eff}(t)
    =& \frac{-i}{\hbar} \left[\hat{H}_\mathrm{0}\,\hat{\rho}_\mathrm{eff}(t)\right] \\ &+ \sum_{j=1}^{n_\mathrm{p}}\frac{\gamma_j}{\hbar} \left(  \hat{a}_j\hat{\rho}_\mathrm{eff}(t)\hat{a}_j^\dagger -\frac{1}{2}\hat{a}_j^\dagger\hat{a}_j\hat{\rho}_\mathrm{eff}(t) -\frac{1}{2}\hat{\rho}_\mathrm{eff}(t)\hat{a}_j^\dagger\hat{a}_j \right),
\label{Eq:Lindblad_Eq}
\end{align}
where $\gamma_j=2\hbar\Gamma_j$ is the optical loss (in some literature, $\gamma_j/\hbar$ is named photon decay rate or cavity dissipation rate\cite{Felicetti2020,Antoniou2020}), and $V_j=\hbar\sqrt{{A_0(\omega_j)F_j\Gamma_j}/{2}}$ in $\hat{H}_0$ [Eq.~(\ref{Eq:H0})] can be regarded as the light-matter coupling strength ($V_j/\hbar$ can be associated with single-particle vacuum Rabi frequency or molecular vacuum Rabi frequency\cite{Herrera2017_2,Herrera2018}). Here, we analytically derive the correspondence between the mQED-based wavefunction approach [Eq.~(\ref{Eq:Dynamic_Equation})] and the cQED-based Lindblad master equation [Eq.~(\ref{Eq:Lindblad_Eq})]. It is noteworthy that the derivation based on the pseudomode approach is not only valid in weak light-matter coupling regimes, but also valid in the strong coupling regimes\cite{Garraway1997,Gonzalez-Tudela2014}.

\section{Numerical Demonstration and Discussion}
\label{Sec:Result}

In the previous section, we have established a theory which connects the mQED-based wavefunction approach [Eq.~(\ref{Eq:Dynamic_Equation})] and the cQED-based Lindblad master equation [Eq.~(\ref{Eq:Lindblad_Eq})] and allows us to estimate the light-matter coupling strength and optical loss. In order to show the use and advantages of our theory, we apply our theory to two representative systems, a spherical silver cavity and a NaCl-coated silver surface. The former is associated with cavity photon, while the latter is associated with plasmon polariton. Via the two systems, our calculations clearly demonstrate that when Purcell factors
behave like Lorentzian functions [Eq.~(\ref{Eq:Lorentz})], our theory can exactly estimate light-matter coupling strength and optical loss. In addition, even if Purcell factors
slightly deviate from Lorentzian functions, our theory can still roughly estimate the two physical quantities and capture the main feature in quantum dynamics.

\subsection{A spherical silver cavity  (cavity photon)}

In the first system, we focus on an excited molecule interacting with a cavity photon mode in a spherical silver cavity, as shown in the inset of Fig.~\ref{fig1}(a), where the molecule is at the center of the spherical silver cavity. In order to estimate the coupling strength and optical loss, we need to calculate the Purcell factor $F_\mathrm{P}(\omega)$ first. The calculation of the Purcell factor requires the material properties of the spherical silver cavity, which includes a vacuum inner layer and a silver outer layer. The spherical silver cavity can be modeled via the dielectric function $\epsilon_\mathrm{r}(\mathbf{r},\omega)$,
\begin{align}
    \epsilon_\mathrm{r}(\mathbf{r},\omega)=
    \begin{cases}
    \epsilon_{\mathrm{r,Vac}} &  \abs{\mathbf{r}}<R \\
    \epsilon_{\mathrm{r,Ag}}(\omega) &  \abs{\mathbf{r}}>R 
    \end{cases} ,
    \label{Eq:Sphere_dielectric}
\end{align}
where $\epsilon_{\mathrm{r,Vac}}$ and $\epsilon_{\mathrm{r,Ag}}(\omega)$ correspond to the dielectric functions of vacuum and silver, respectively. Furthermore, we also need to evaluate the imaginary part of the dyadic Green's function 
$\mathrm{Im}\left\{\overline{\overline{\mathbf{G}}}(\mathbf{r}_\mathrm{M},\mathbf{r}_\mathrm{M},\omega)\right\}$ of this spherical cavity structure as\cite{Le-WeiLi1994,Scheel1999}
{\small
\begin{align}
    &\left. \mathrm{Im}\left\{\overline{\overline{\mathbf{G}}}(\mathbf{r}_\mathrm{M},\mathbf{r}_\mathrm{M},\omega)\right\} \right\vert_{\mathbf{r}_\mathrm{M}=0} = \mathrm{Im} \left\{ \frac{i\omega}{6\pi c } \left(1+C_1^\mathrm{N}(\omega)\right) \begin{bmatrix} 1 & 0 & 0 \\ 0 & 1 & 0 \\ 0 & 0 & 1 \end{bmatrix} \right\} ,
    \label{Eq:Dyadic_Green_Spherical}
\end{align} }
where 
{\small
\begin{align}
\nonumber
    C_1^\mathrm{N}(\omega) &= \frac{\left(i+z(n+1)-iz^2n-\frac{z^3n^2}{n+1}\right)e^{iz}}{\mathrm{sin}z -z\left(\mathrm{cos}z-in\mathrm{sin}z\right)+iz^2n\mathrm{cos}z-\frac{z^3\left(\mathrm{cos}z-in\mathrm{sin}z\right)n^2}{n^2-1}}, \\
\nonumber
    z &= \frac{R\omega}{c}, \\
\nonumber
    n &= \sqrt{\epsilon_{\mathrm{r,Ag}}(\omega)},
\end{align}
}where $R$ is the radius of the cavity, and is set to be $200\,\mathrm{nm}$. The data of the dielectric function $\epsilon_{\mathrm{r,Ag}}(\omega)$ is adopted from the work of Johnson and Christy\cite{Johnson1972}. Based on Eq.~(\ref{Eq:Dyadic_Green_Spherical}), we can calculate the Purcell factor $F_\mathrm{P}(\omega)=A_\mathrm{FG}(\omega)/A_0(\omega)$ via the spontaneous emission rate of the molecule at the center of the spherical silver cavity $A_\mathrm{FG}(\omega)$ in Eq.~(\ref{Eq:Media_Decay_Rate}) and the spontaneous emission rate of the molecule in vacuum $A_0(\omega)$ in Eq.~(\ref{Eq:Vacuum_Decay_Rate}). The Purcell factor (a green solid line) is shown in Fig.~\ref{fig1}(a).

Obviously, the Purcell factor in Fig.~\ref{fig1}(a) resembles a Lorentzian function so that we can use Eq.~(\ref{Eq:Lorentz}) to fit the green line with $n_\mathrm{p}=1$, and obtain $\hbar \omega_1\approx 2.3045\,\mathrm{eV}$ and $F_1\approx 211.47$ by the peak position and the peak height of the calculated Purcell factor. $\hbar \Gamma_1\approx 5.2588\,\mathrm{meV}$ (i.e., $ \gamma_1 \approx 10.518\,\mathrm{meV}$) is the only parameter fitted by the least squares method, and the fitted curve is plotted with a yellow dashed line as shown in Fig.~\ref{fig1}(a). In order to calculate the light-matter coupling strength $V_1=\hbar\sqrt{{A_0(\omega_1)F_1\Gamma_1}/{2}}$, the spontaneous emission rate $A_0(\omega_1)$ and the related molecular properties are needed, which are chosen to be $\hbar\omega_\mathrm{eg}=\hbar\omega_1 = 2.3045 \,\mathrm{eV}$, $\abs{\boldsymbol{\mu}}=15\,\mathrm{D}$ (corresponds to $A_0(\omega_1)\approx 4.5312\times 10^8\,\mathrm{s^{-1}}$ or $\hbar A_0(\omega_1)\approx 2.9825\times 10^{-4}\,\mathrm{meV}$), $\omega_{\mathrm{vib},1}=1209.8\,\mathrm{cm^{-1}}\approx 0.15000\,\mathrm{eV}$, $\omega_{\mathrm{vib},2}=80.655\,\mathrm{cm^{-1}}\approx0.01000\,\mathrm{eV}$, $S_1=0.5$, and $S_2=0.1$. Therefore, the coupling strength can be calculated as $V_1=\hbar\sqrt{{ A_0(\omega_1)F_1\Gamma_1}/{2}}=\sqrt{{\hbar A_0(\omega_1)F_1\hbar\Gamma_1}/{2}}\approx 0.40722\,\mathrm{meV}$.

To numerically demonstrate the consistency between the mQED-based wavefunction approach and the cQED-based Lindblad master equation, we calculate the population dynamics $P^{\mathrm{e},\left\{0\right\}}_{\left\{M\right\}_\mathrm{vib}}(t)=\abs{C^{\mathrm{e},\left\{0\right\}}_{\left\{M\right\}_\mathrm{vib}}(t)}^2$ of the molecule based on Eq.~(\ref{Eq:Dynamic_Equation}) and Eq.~(\ref{Eq:Lindblad_Eq}), in which $P^{\mathrm{e},\left\{0\right\}}_{\left\{M\right\}_\mathrm{vib}}(t)$ can be specified as $P^{\mathrm{e},\left\{0\right\}}_{M_1,M_2}(t)$, where the subscript $M_{1(2)}$ denotes the quantum of the first (second) vibrational mode. We set the initial condition $C^{\mathrm{e},\left\{0\right\}}_{0,0}(0)=1$ and use the flat continuum approximation in Eq.~(\ref{Eq:Dynamic_Equation}) in order to compare it with the dynamics derived from Eq.~(\ref{Eq:Lindblad_Eq}). As shown in Figs.~\ref{fig1}(b), \ref{fig1}(c), and \ref{fig1}(d), the population dynamics $P^{\mathrm{e},\left\{0\right\}}_{M_1,M_2}(t)$ calculated by the mQED-based wavefunction approach [Eq.~(\ref{Eq:Dynamic_Equation})] with the exact Purcell factor (green solid line) perfectly matches the population dynamics calculated with the Lorentz-shape Purcell factor (yellow dashed line) because the exact Purcell factor behaves like a Lorentzian function, as shown in Fig.~\ref{fig1}(a). Moreover, the population dynamics obtained from the cQED-based Lindblad master equation (red solid line) also almost coincides with the population dynamics calculated by the mQED-based wavefunction approach (yellow dashed line), which indicates that the mapping from Eq.~(\ref{Eq:Dynamic_Equation}) to Eq.~(\ref{Eq:Lindblad_Eq}) is exact. That is, one can apply the cQED-based Lindblad master equation to quantitatively investigate the quantum dynamics in a dispersive and lossy cavity, and this method is exactly the same as our original mQED-based wavefunction theory.

Note that the coupling strength between the molecule and the cavity photon $V_1\approx 0.40722\,\mathrm{meV}$ is much smaller than $\hbar\Gamma_1 \approx 5.2588\,\mathrm{meV}$, leading to a Markovian exponential decay of $P^{\mathrm{e},\left\{0\right\}}_{0,0}(t)$, and the decay constant can be described by the formula $\prod_{u=1}^2\abs{\braket{0(0)}{0(S_u)}}^2 F_1 A_0(\omega_1)$ in our previous study\cite{Wang2020}. Incidentally, we also find that the oscillation frequencies of $P_{1,0}^{\mathrm{e},\left\{0\right\}}(t)$ Fig.~\ref{fig1}(c) and $P_{0,1}^{\mathrm{e},\left\{0\right\}}(t)$ in Fig.~\ref{fig1}(d) correspond to  $\omega_{\mathrm{vib},1}$ and $\omega_{\mathrm{vib},2}$, respectively. Note that the envelopes of $P_{1,0}^{\mathrm{e},\left\{0\right\}}(t)$ and $P_{0,1}^{\mathrm{e},\left\{0\right\}}(t)$ do not exponentially decay because we plot $P_{1,0}^{\mathrm{e},\left\{0\right\}}(t)$ and $P_{0,1}^{\mathrm{e},\left\{0\right\}}(t)$ in a short time range. In a long time range, their decay rates are almost the same as that of $P_{0,0}^{\mathrm{e},\left\{0\right\}}(t)$.

\begin{widetext}
    \begin{center}
            \begin{figure} [t!]
            \includegraphics[width=1\textwidth]{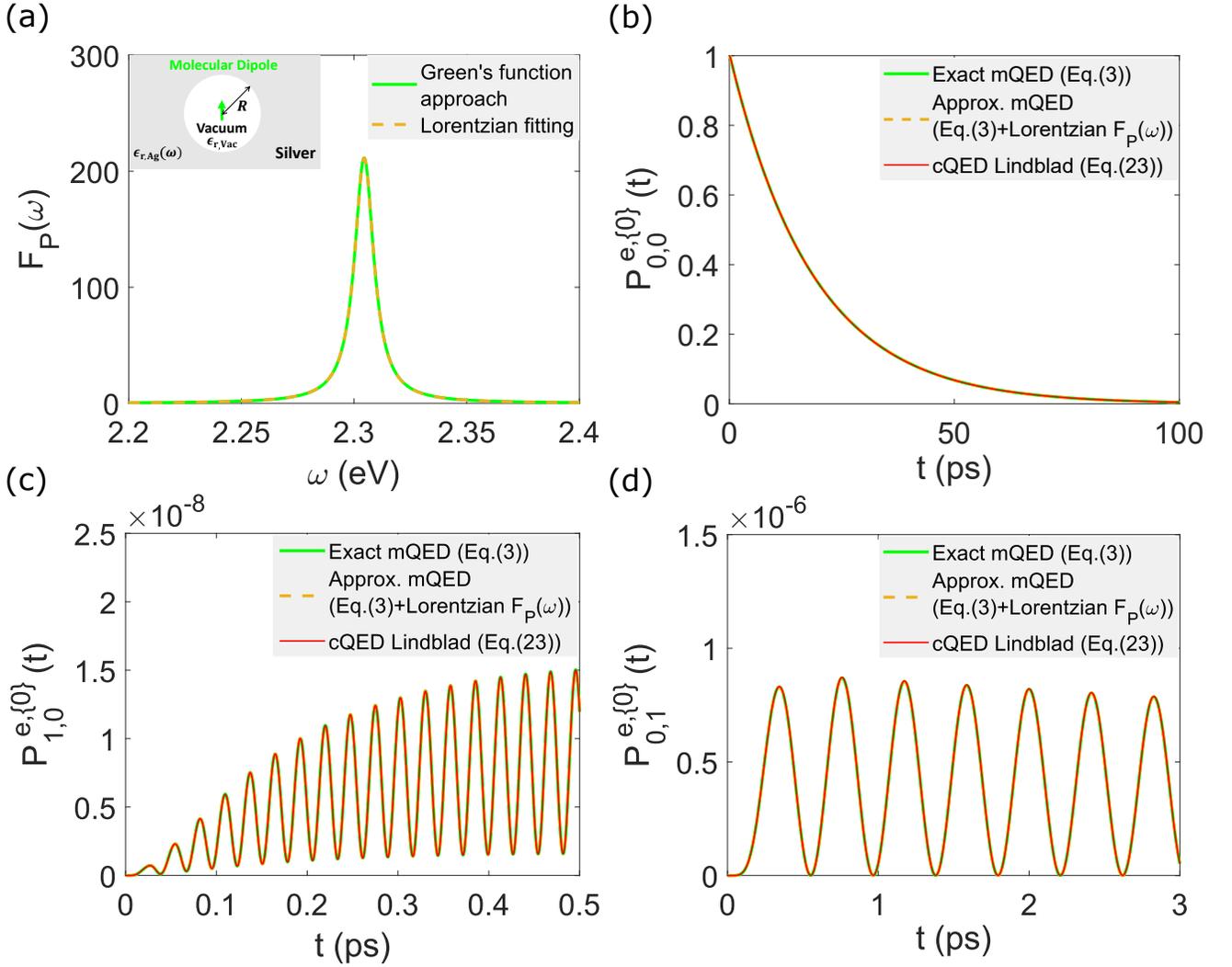}
            \caption{Purcell factor of a molecule in a spherical silver cavity and the corresponding population dynamics based on Eq.~(\ref{Eq:Dynamic_Equation}) and Eq.~(\ref{Eq:Lindblad_Eq}). (a) Purcell factor derived from Eqs.~(\ref{Eq:Media_Decay_Rate}), (\ref{Eq:Vacuum_Decay_Rate}), and (\ref{Eq:Dyadic_Green_Spherical}). The computational result indicates that the Purcell factor can be fitted by a Lorentzian function. Inset: Sketch of a molecular emitter at the center of a spherical silver cavity. (b) Population dynamics of $P_{0,0}^{\mathrm{e},\left\{0\right\}}(t)$ (the subscript ${0,0}$ denotes zero quanta for all vibrational modes), (c) Population dynamics of $P_{1,0}^{\mathrm{e},\left\{0\right\}}(t)$ (the subscript ${1,0}$ denotes the first vibrational mode with 1 quantum and the second vibrational mode with 0 quanta), and (d) Population dynamics of $P_{0,1}^{\mathrm{e},\left\{0\right\}}(t)$ (the subscript ${0,1}$ denotes the second vibrational mode with 1 quantum and the first vibrational mode with 0 quanta). }
            \label{fig1}
        \end{figure}
    \end{center}
\end{widetext}

\subsection{A NaCl-coated silver surface (plasmon polariton)}

In the second system, we focus on an excited molecule interacting with a plasmon polariton mode on a NaCl-coated silver surface, as shown in the inset of Fig.~\ref{fig2}(a), where the distance between the molecule and the NaCl layer is $h_1=0.1\,\mathrm{nm}$, and the thickness of the NaCl layer is $h_2=0.9\,\mathrm{nm}$ (that is, the distance between the molecule and the silver surface $d=h_1+h_2$ is $1.0\,\mathrm{nm}$). We model this system via the following dielectric function $\epsilon_\mathrm{r}(\mathbf{r},\omega)$,
\begin{align}
    \epsilon_\mathrm{r}(\mathbf{r},\omega)=
    \begin{cases}
    \epsilon_{\mathrm{r,Vac}} &  z>0 \\
    \epsilon_{\mathrm{r,NaCl}}(\omega) &  0>z>-h_2 \\
    \epsilon_{\mathrm{r,Ag}}(\omega) & z<-h_2
    \end{cases},
    \label{Eq:dielectric}
\end{align}
where $\epsilon_{\mathrm{r,Vac}}$ and $\epsilon_{\mathrm{r,Ag}}(\omega)$ are the same as in the first system. $\epsilon_{\mathrm{r,NaCl}}(\omega)$ corresponds to the dielectric function of NaCl and is adopted from the work of Li\cite{Li1976}. The imaginary part of the dyadic Green's function $\mathrm{Im}\left\{\overline{\overline{\mathbf{G}}}(\mathbf{r}_\mathrm{M},\mathbf{r}_\mathrm{M},\omega)\right\}$ required for the calculation of the Purcell factor $F_\mathrm{P}(\omega)$ can be obtained by integrating the
reciprocal space (the details can be found in Appendix~\ref{Appendix_B}). Combining the dyadic Green's function, Eq.~(\ref{Eq:Media_Decay_Rate}), and Eq.~(\ref{Eq:Vacuum_Decay_Rate}), we can calculate the exact Purcell factor of a molecule emitter whose transition dipole moment is vertical to the surface, and the exact Purcell factor (a green solid line) is shown in Fig.~\ref{fig2}(a).

\begin{widetext}
    \begin{center}
            \begin{figure}[t!]
            \includegraphics[width=1\textwidth]{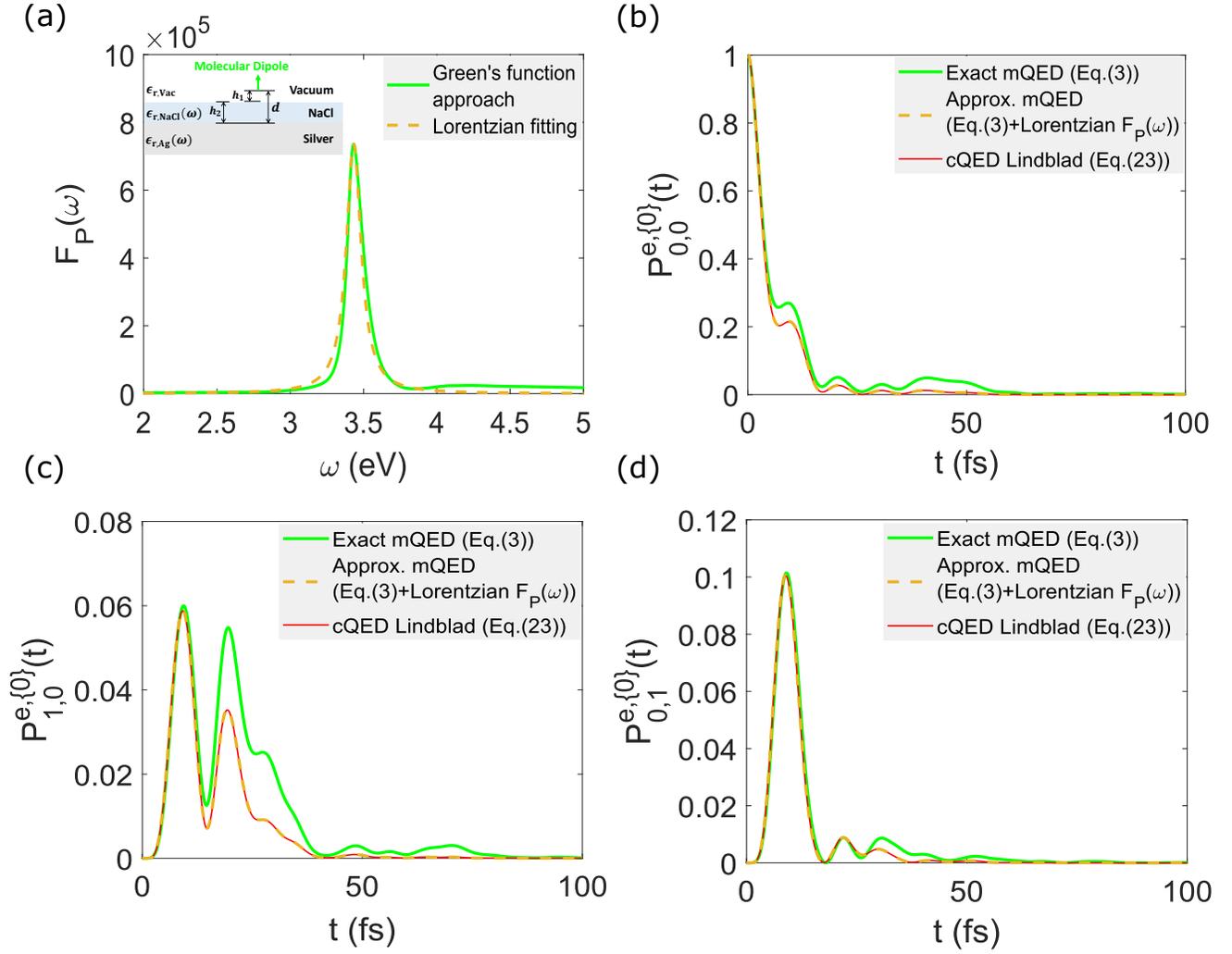}
            \caption{Purcell factor of a molecule on a NaCl-coated silver surface and the corresponding population dynamics based on Eq.~(\ref{Eq:Dynamic_Equation}) and Eq.~(\ref{Eq:Lindblad_Eq}). (a) Purcell factor derived from Eqs.~(\ref{Eq:Media_Decay_Rate}), (\ref{Eq:Vacuum_Decay_Rate}), and (\ref{Eq:Dyadic_Green_Plane}). Inset: Sketch of a molecular emitter on a NaCl-coated silver surface. (b) Population dynamics of $P_{0,0}^{\mathrm{e},\left\{0\right\}}(t)$, (c) Population dynamics of $P_{1,0}^{\mathrm{e},\left\{0\right\}}(t)$, and (d) Population dynamics of $P_{0,1}^{\mathrm{e},\left\{0\right\}}(t)$. }
            \label{fig2}
        \end{figure}
    \end{center}
\end{widetext}

Following the same procedure in Sec.~\ref{Sec:Result}A, we obtain a fitted Lorentz-shape Purcell factor (yellow dashed line) with $\hbar \omega_1\approx 3.4325\,\mathrm{eV}$, $F_1\approx 7.3585\times 10^{5}$, and $\hbar \Gamma_1 \approx 61.750\,\mathrm{meV}$ (i.e., $ \gamma_1 \approx 123.50\,\mathrm{meV}$), as shown in Fig.~\ref{fig2}(a). The spontaneous emission rate $A_0(\omega_1)$ and the related molecular properties are chosen to be $\hbar\omega_\mathrm{eg}=\hbar\omega_1 = 3.4325 \,\mathrm{eV}$, $\abs{\boldsymbol{\mu}}=18\,\mathrm{D}$ (corresponds to $A_0(\omega_1)\approx 2.1562\times 10^9\,\mathrm{s^{-1}}$ or $\hbar A_0(\omega_1)\approx 1.4192\times 10^{-3}\,\mathrm{meV}$), $\omega_{\mathrm{vib},1}=1000.0\,\mathrm{cm^{-1}}\approx 0.12398\,\mathrm{eV}$, $\omega_{\mathrm{vib},2}=1700.0\,\mathrm{cm^{-1}}\approx 0.21077\,\mathrm{eV}$, $S_1=0.8$, and $S_2=0.6$. Therefore, the coupling strength can be calculated as $V=\hbar\sqrt{{ A_0(\omega_1)F_1\Gamma_1}/{2}}=\sqrt{{\hbar A_0(\omega_1)F_1\hbar\Gamma_1}/{2}}\approx 179.56 \,\mathrm{meV}$.

Similarly, we calculate the population dynamics $P^{\mathrm{e},\left\{0\right\}}_{\left\{M\right\}_\mathrm{vib}}(t)$ with the same premises in Sec.~\ref{Sec:Result}A. From Figs.~\ref{fig2}(b), \ref{fig2}(c), and \ref{fig2}(d), we notice slight differences between the population dynamics $P^{\mathrm{e},\left\{0\right\}}_{M_1,M_2}(t)$ calculated by the mQED-based  wavefunction approach [Eq.~(\ref{Eq:Dynamic_Equation})] with the exact Purcell factor (green solid line) and with the Lorentz-shape Purcell factor (yellow dashed line) due to the deviation of the exact Purcell factor from a perfect Lorentzian function, as shown in Fig.~\ref{fig2}(a). On the other hand, the population dynamics obtained from the cQED-based Lindblad master equation (red solid line) still almost coincides with the population dynamics calculated by the mQED-based wavefunction approach with the Lorentz-shape Purcell factor (yellow dashed line), which indicates that the mapping from Eq.~(\ref{Eq:Dynamic_Equation}) to Eq.~(\ref{Eq:Lindblad_Eq}) is still exact. The similarity between the red solid line and the green solid line indicates that one can apply the cQED-based Lindblad master equation to qualitatively investigate the quantum dynamics in a plasmonic system. Incidentally, the coupling strength between the molecule and the plasmon polariton $V_1\approx 179.56\,\mathrm{meV}$ is much larger than $\hbar\Gamma_1 \approx 61.750\,\mathrm{meV}$, leading to a non-Markovian Rabi oscillation of population $P^{\mathrm{e},\left\{0\right\}}_{0,0}(t)$\cite{Wang2020,Wang2020a}.

From these two systems, we not only show how to calculate the light-matter coupling strength and optical loss, but also numerically demonstrate the correspondence between the mQED-based wavefunction approach [Eq.~(\ref{Eq:Dynamic_Equation})] and the cQED-based Lindblad master equation [Eq.~(\ref{Eq:Lindblad_Eq})]. Considering that the integro-differential equation [Eq.~(\ref{Eq:Dynamic_Equation})] is computationally expensive, the use of Eq.~(\ref{Eq:Lindblad_Eq}) instead of Eq.~(\ref{Eq:Dynamic_Equation}) significantly reduces the computational cost and provides a clear physical insight in light-matter interactions in complicated dielectric environments.

\section{Conclusions}
\label{Sec:Conclusion}

In this study, we successfully bridged macroscopic quantum electrodynamics and cavity quantum electrodynamics through the pseudomode approach. On the basis of the Lorentz-shape Purcell factor [Eq.~(\ref{Eq:Lorentz})] and the flat continuum approximation, we rigorously proved that our mQED-based wavefunction theory [Eq.~(\ref{Eq:Dynamic_Equation})] can be mapped into the cQED-based Lindblad master equation [Eq.~(\ref{Eq:Lindblad_Eq})]. 
Moreover, from the mapping, the light-matter coupling strength $V$ and the optical loss $\gamma$ used in many cQED models\cite{Herrera2017_2,Herrera2018,Felicetti2020,Antoniou2020} can be calculated via the two equations,
\begin{align}
    \nonumber
    V&=\hbar\sqrt{{A_0(\omega)F\Gamma}/{2}}, \\
    \gamma&=2\hbar\Gamma,\nonumber
\end{align}
where $\omega$, $F$ and $\Gamma$ are the peak position, the peak height, and the half width at half maximum of a Lorentz-shape Purcell factor, respectively; $A_0(\omega)$ is the molecular spontaneous emission rate in vacuum. 
To demonstrate the advantage of our theory, we investigated the quantum dynamics of molecular fluorescence in two representative systems. In the first system, we studied an excited molecule weakly coupled with a cavity photon mode ($V<\gamma/2$), and the dynamics described by $V$ and $\gamma$ from our present theory exactly coincide with the dynamics derived from our previous mQED-based approach. In the second system, we studied an excited molecule strongly coupled with a plasmon polariton mode ($V>\gamma/2$). Although in this case the Purcell factor slightly deviates from a perfect Lorentzian function, $V$ and $\gamma$ can still be roughly estimated by our theory. Here, we would like to emphasize that the light-matter coupling strength given by our theory can be applied to experiments. By using $\hbar\omega\approx 1.86\,\mathrm{eV}$, $\hbar A_0(\omega)\approx 1.0136\,\times 10^{-5}\,\mathrm{meV}$ (which corresponds to $A_0(\omega)\approx 1.54\times 10^7\,\mathrm{s^{-1}}$), $F \approx 3.5\times 10^6$, and $\hbar\Gamma\approx 84 \, \mathrm{meV}$ from Cavendish Laboratory\cite{Chikkaraddy2016}, one can quickly estimates the coupling strengths $V\approx 39 \, \mathrm{meV}$ (i.e., $2V\approx 78 \, \mathrm{meV}$)\cite{Wang2019}, which is in agreement with the experimentally observed Rabi splitting $2V\approx 80$ -- $95 \, \mathrm{meV}$.  

Although we have clearly demonstrated a mapping from our mQED-based wavefunction approach to the cQED-based Lindblad master equation, several issues still remain to be addressed. First, the current theory is restricted to the condition that the Purcell factor behaves like Lorentzian functions. For a non-Lorentzian Purcell factor, the mapping may become complicated, and the resulting cQED-based dynamical equations may involve multiple pseudomodes which are coupled with each other\cite{Dalton2001,Pleasance2020}. Second, for the deep-strong-coupling regime ($V>\hbar\omega_\mathrm{eg}$)\cite{FriskKockum2019}, bound states can be formed of molecular excitation and photons, the approximation in Eq.~(\ref{Eq:Extend_lower}) may become inappropriate\cite{Yang2017,Wen2020}. These issues will be further explored in our future work. Macroscopic quantum electrodynamics is a powerful methodology for exploring light-matter interactions in complicated dielectric environments, and we have successfully established general theories of resonance energy transfer\cite{Ding2017,Hsu2017,Hsu_2018,Lee2020} and molecular fluorescence\cite{Wang2019,Wang2020,Wang2020a} in the framework of mQED. We hope that this study could motivate more experimental and theoretical investigations into molecules coupled with photonic modes in complicated dielectric environments.

\begin{acknowledgments}
Wang, Chuang and Hsu thanks Yu-Chen Wei for manuscript reading. Hsu thanks Academia Sinica and the Ministry of Science and Technology of Taiwan (MOST 109-2113-M-001-021-) for the financial support.
\end{acknowledgments}

\section*{data availability}
The data that support the findings of this study are available from the corresponding author upon reasonable request.

\begin{appendix}

\begin{widetext}

\section{Eqs.~(\ref{Eq:14}) and (\ref{Eq:15}) derived from Eqs.~(\ref{Eq:16}) and (\ref{Eq:Effective_Hamiltonian})}
\label{Appendix_A}
Substituting the wavefunction ansatz Eq.~(\ref{Eq:16}) and the effective Hamiltonian Eq.~(\ref{Eq:Effective_Hamiltonian}) into the Schr\"{o}dinger equation $i\hbar\frac{d}{dt}\ket{\varPhi(t)}=\hat{H}_\mathrm{eff}\ket{\varPhi(t)}$, the left-hand side turns to be
{\small
\begin{align}
\nonumber
    i\hbar\frac{d}{dt}\ket{\varPhi(t)}=& \sum_{\left\{M\right\}_\mathrm{vib}=0}^\infty \left\{ i\hbar\dot{C}_{\left\{M\right\}_\mathrm{vib}}^\mathrm{e,\left\{0\right\}}(t) +\hbar\left(\omega_\mathrm{eg}+\sum_{u=1}^{n_\mathrm{vib}}\omega_{\mathrm{vib},u}M_u\right){C}_{\left\{M\right\}_\mathrm{vib}}^\mathrm{e,\left\{0\right\}}(t) \right\} e^{-i(\omega_\mathrm{eg}+\sum_{u=1}^{n_\mathrm{vib}}\omega_{\mathrm{vib},u}M_u)t} \ket{\mathrm{e},\left\{0\right\}_\mathrm{p}} \prod_{u=1}^{n_\mathrm{vib}}\ket{M_u(0)}\\
    &+\sum_{j=1}^{n_\mathrm{p}}\sum_{\left\{M'\right\}_\mathrm{vib}=0}^\infty \left\{i\hbar\dot{B}_{\left\{M'\right\}_\mathrm{vib}}^{\mathrm{g},\left\{1_j\right\}_\mathrm{p}} (t)+\hbar\left(\omega_j-i\Gamma_j+\sum_{u=1}^{n_\mathrm{vib}}\omega_{\mathrm{vib},u}M'_u\right){B}_{\left\{M'\right\}_\mathrm{vib}}^{\mathrm{g},\left\{1_j\right\}_\mathrm{p}}(t) \right\} e^{-i(\omega_j-i\Gamma_j+\sum_{u=1}^{n_\mathrm{vib}}\omega_{\mathrm{vib},u}M'_u)t} \ket{\mathrm{g},\left\{1\right\}_\mathrm{p}} \prod_{u=1}^{n_\mathrm{vib}}\ket{M'_u(S_u)},
\end{align}
}and the right-hand side becomes
{\small
\begin{align}
\nonumber
   &\hat{H}_\mathrm{eff}\ket{\varPhi(t)}  \\
\nonumber
    &= \sum_{\left\{M\right\}_\mathrm{vib}=0}^\infty \hbar\left( \omega_{\mathrm{eg}} + \sum_{u=1}^{n_\mathrm{vib}}  \omega_{\mathrm{vib},u} M_u \right)C_{\left\{M\right\}_\mathrm{vib}}^\mathrm{e,\left\{0\right\}}(t) e^{-i(\omega_\mathrm{eg}+\sum_{u=1}^{n_\mathrm{vib}}\omega_{\mathrm{vib},u}M_u)t}  \ket{\mathrm{e},\left\{0\right\}_\mathrm{p}} \prod_{u=1}^{n_\mathrm{vib}}\ket{M_u(0)}\\
\nonumber
    &\quad -\sum_{j=1}^{n_\mathrm{p}}\sum_{\left\{M\right\}_\mathrm{vib}=0}^\infty V_j C_{\left\{M\right\}_\mathrm{vib}}^\mathrm{e,\left\{0\right\}}(t) e^{-i(\omega_\mathrm{eg}+\sum_{u=1}^{n_\mathrm{vib}}\omega_{\mathrm{vib},u}M_u)t}  \ket{\mathrm{g},\left\{1_j\right\}_\mathrm{p}} \prod_{u=1}^{n_\mathrm{vib}}\ket{M_u(0)}- \sum_{j=1}^{n_\mathrm{p}}\sum_{\left\{M'\right\}_\mathrm{vib}=0}^\infty V_j B_{\left\{M'\right\}_\mathrm{vib}}^{\mathrm{g},\left\{1_j\right\}_\mathrm{p}}(t) e^{-i(\omega_j-i\Gamma_j+\sum_{u=1}^{n_\mathrm{vib}}\omega_{\mathrm{vib},u}M'_u)t} \ket{\mathrm{e},\left\{0\right\}_\mathrm{p}} \prod_{u=1}^{n_\mathrm{vib}}\ket{M'_u(S_u)}\\
    &\quad +\sum_{j=1}^{n_\mathrm{p}}\sum_{\left\{M'\right\}_\mathrm{vib}=0}^\infty \hbar\left(\omega_j-i\Gamma_j + \sum_{u=1}^{n_\mathrm{vib}}\omega_{\mathrm{vib},u}M_u'  \right) B_{\left\{M'\right\}_\mathrm{vib}}^{\mathrm{g},\left\{1_j\right\}_\mathrm{p}}(t) e^{-i(\omega_j-i\Gamma_j+\sum_{u=1}^{n_\mathrm{vib}}\omega_{\mathrm{vib},u}M'_u)t} \ket{\mathrm{g},\left\{1\right\}_\mathrm{p}} \prod_{u=1}^{n_\mathrm{vib}}\ket{M'_u(S_u)}.
\label{Eq:A2}
\end{align}}

Comparing the left-hand side and the right-hand side of the Schr\"{o}dinger equation, we obtain
{\small
\begin{align}
\nonumber
    &\sum_{\left\{M\right\}_\mathrm{vib}=0}^\infty \left\{ i\hbar\dot{C}_{\left\{M\right\}_\mathrm{vib}}^\mathrm{e,\left\{0\right\}}(t) \right\} e^{-i(\omega_\mathrm{eg}+\sum_{u=1}^{n_\mathrm{vib}}\omega_{\mathrm{vib},u}M_u)t} \ket{\mathrm{e},\left\{0\right\}_\mathrm{p}} \prod_{u=1}^{n_\mathrm{vib}}\ket{M_u(0)}+
    \sum_{j=1}^{n_\mathrm{p}}\sum_{\left\{M'\right\}_\mathrm{vib}=0}^\infty \left\{i\hbar\dot{B}_{\left\{M'\right\}_\mathrm{vib}}^{\mathrm{g},\left\{1_j\right\}_\mathrm{p}}(t) \right\} e^{-i(\omega_j-i\Gamma_j+\sum_{u=1}^{n_\mathrm{vib}}\omega_{\mathrm{vib},u}M'_u)t} \ket{\mathrm{g},\left\{1_j\right\}_\mathrm{p}} \prod_{u=1}^{n_\mathrm{vib}}\ket{M'_u(S_u)}\\
    &=-\sum_{j=1}^{n_\mathrm{p}}\sum_{\left\{M\right\}_\mathrm{vib}=0}^\infty V_j C_{\left\{M\right\}_\mathrm{vib}}^\mathrm{e,\left\{0\right\}}(t) e^{-i(\omega_\mathrm{eg}+\sum_{u=1}^{n_\mathrm{vib}}\omega_{\mathrm{vib},u}M_u)t}  \ket{\mathrm{g},\left\{1_j\right\}_\mathrm{p}} \prod_{u=1}^{n_\mathrm{vib}}\ket{M_u(0)}- \sum_{j=1}^{n_\mathrm{p}}\sum_{\left\{M'\right\}_\mathrm{vib}=0}^\infty V_j B_{\left\{M'\right\}_\mathrm{vib}}^{\mathrm{g},\left\{1_j\right\}_\mathrm{p}}(t) e^{-i(\omega_j-i\Gamma_j+\sum_{u=1}^{n_\mathrm{vib}}\omega_{\mathrm{vib},u}M'_u)t} \ket{\mathrm{e},\left\{0\right\}_\mathrm{p}} \prod_{u=1}^{n_\mathrm{vib}}\ket{M'_u(S_u)} .
\label{Eq:A3}
\end{align}}

Let $\bra{\mathrm{e},\left\{0\right\}_\mathrm{p}}\prod_{u=1}^{n_\mathrm{vib}}\bra{K_u(0)}$ act on Eq.~(\ref{Eq:A3}), we obtain the following equation:
\begin{align}
    i\hbar\dot{C}_{\left\{K\right\}_\mathrm{vib}}^\mathrm{e,\left\{0\right\}}(t) e^{-i(\omega_\mathrm{eg}+\sum_{u=1}^{n_\mathrm{vib}}\omega_{\mathrm{vib},u}K_u)t} = -\sum_{j=1}^{n_\mathrm{p}}\sum_{\left\{M'\right\}_\mathrm{vib}=0}^\infty V_j B_{\left\{M'\right\}_\mathrm{vib}}^{\mathrm{g},\left\{1_j\right\}_\mathrm{p}}(t) e^{-i(\omega_j-i\Gamma_j+\sum_{u=1}^{n_\mathrm{vib}}\omega_{\mathrm{vib},u}M'_u)t}  \prod_{u=1}^{n_\mathrm{vib}}\braket{K_u(0)}{M'_u(S_u)} .
\label{Eq:A4}
\end{align}

Because of $V_j=\hbar\sqrt{A_0(\omega_j)F_j\Gamma_j/2}$, we can express Eq.~(\ref{Eq:A4}) as [recall Eq.~(\ref{Eq:14})]:
\begin{align}
    \frac{d C_{\left\{K\right\}_\mathrm{vib}}^\mathrm{e,\left\{0\right\}}(t)}{dt} =  i \sum_{j=1}^{n_\mathrm{p}} \sum_{\left\{M'\right\}_\mathrm{vib}=0}^\infty \left\{ \sqrt{\frac{A_0(\omega_j)F_j\Gamma_j}{2}}  e^{-i(\omega_j-\omega_\mathrm{eg}-i\Gamma_j)t} \right\} \left\{ \prod_{u=1}^{n_\mathrm{vib}}\braket{K_u(0)}{M'_u(S_u)} e^{i\omega_{\mathrm{vib},u}(K_u-M'_u)t} \right\} B_{\left\{M'\right\}_\mathrm{vib}}^{\mathrm{g},\left\{1_j\right\}_\mathrm{p}}(t).
\end{align}

Similarly, by letting $\bra{\mathrm{g},\left\{1_j\right\}_\mathrm{p}}\prod_{u=1}^{n_\mathrm{vib}}\bra{M_u'(S_u)}$ act on Eq.~(\ref{Eq:A3}), we obtain another differential equation [recall Eq.~(\ref{Eq:15})]:
\begin{align}
    \frac{dB_{\left\{M'\right\}_\mathrm{vib}}^{\mathrm{g},\left\{1_j\right\}_\mathrm{p}}(t)}{dt} = i \sum_{\left\{M\right\}_\mathrm{vib}=0}^\infty\left\{ \sqrt{\frac{A_0(\omega_j)F_j\Gamma_j}{2}}   e^{-i(\omega_\mathrm{eg}-\omega_j+i\Gamma_j)t} \right\}\left\{ \prod_{u=1}^{n_\mathrm{vib}}\braket{M'_u(S_u)}{M_u(0)} e^{i\omega_{\mathrm{vib},u}(M'_u-M_u)t} \right\} C_{\left\{M\right\}_\mathrm{vib}}^\mathrm{e,\left\{0\right\}}(t).
\end{align}
\end{widetext}

\section{Dyadic Green's function of NaCl-coated silver structure}
\label{Appendix_B}
The imaginary part of the dyadic Green's function $\mathrm{Im}\left\{\overline{\overline{\mathbf{G}}}(\mathbf{r}_\mathrm{M},\mathbf{r}_\mathrm{M},\omega)\right\}$ of the NaCl-coated silver system can be expressed as\cite{Tomas1995,Nikitin2013,Hsu_2018}
{\small
\begin{align}
\nonumber
    \mathrm{Im}\left\{\overline{\overline{\mathbf{G}}}(\mathbf{r}_\mathrm{M},\mathbf{r}_\mathrm{M},\omega)\right\} = & \mathrm{Im}\left\{\overline{\overline{\mathbf{G}}}_0(\mathbf{r}_\mathrm{M},\mathbf{r}_\mathrm{M},\omega)\right\} + \mathrm{Im}\left\{\overline{\overline{\mathbf{G}}}_\mathrm{s}(\mathbf{r}_\mathrm{M},\mathbf{r}_\mathrm{M},\omega)\right\}\\& + \mathrm{Im}\left\{\overline{\overline{\mathbf{G}}}_\mathrm{p}(\mathbf{r}_\mathrm{M},\mathbf{r}_\mathrm{M},\omega)\right\},
\label{Eq:Dyadic_Green_Plane}
\end{align}
}where $\mathrm{Im}\left\{\overline{\overline{\mathbf{G}}}_0(\mathbf{r}_\mathrm{M},\mathbf{r}_\mathrm{M},\omega)\right\}$, $\mathrm{Im}\left\{\overline{\overline{\mathbf{G}}}_\mathrm{s}(\mathbf{r}_\mathrm{M},\mathbf{r}_\mathrm{M},\omega)\right\}$ and $\mathrm{Im}\left\{\overline{\overline{\mathbf{G}}}_\mathrm{p}(\mathbf{r}_\mathrm{M},\mathbf{r}_\mathrm{M},\omega)\right\}$ are the imaginary parts of the free-space dyadic Green's function, s-polarized dyadic Green's function and p-polarized dyadic Green's function, respectively. For the NaCl-coated silver system, we calculate these imaginary parts based on the following equations \cite{Dung2002,Hsu_2018,Wang2020},
{\small  
\begin{align}
    &\mathrm{Im}\left\{\overline{\overline{\mathbf{G}}}_0(\mathbf{r}_\mathrm{M},\mathbf{r}_\mathrm{M},\omega)\right\} = \frac{\omega}{6\pi c }\mathbf{\overline{\overline{I}}}_3 = \frac{\omega}{6\pi c }\begin{bmatrix} 1 & 0 & 0 \\ 0 & 1 & 0 \\ 0 & 0 & 1 \end{bmatrix} ,
\end{align}
\begin{align}
\nonumber
    &\mathrm{Im}\left\{\overline{\overline{\bm{G}}}_\mathrm{s}(\mathbf{r}_\mathrm{M},\mathbf{r}_\mathrm{M},\omega) \right\} \\
    &= \mathrm{Im} \left\{ \int_0^\infty dq \frac{iqR_\mathrm{s}(q,\omega)}{8\pi K_{z,\mathrm{Vac}}(q,\omega)}    \begin{bmatrix} 1 & 0 & 0 \\ 0 & 1 & 0 \\ 0 & 0 & 0 \end{bmatrix} e^{2iK_{z,\mathrm{Vac}}(q,\omega)h_1} \right\},  \\
\nonumber
    &\mathrm{Im}\left\{ \overline{\overline{\bm{G}}}_\mathrm{p}(\mathbf{r}_\mathrm{M},\mathbf{r}_\mathrm{M},\omega) \right\}= \mathrm{Im} \left\{ \int_0^\infty dq \frac{-iqK_{z,\mathrm{Vac}}(q,\omega)R_\mathrm{p}(q,\omega) }{8\pi (\omega/c)^2} \right.
     \\
    & \qquad \qquad \qquad \qquad \quad \left. \times   \begin{bmatrix} 1 & 0 & 0 \\ 0 & 1 & 0 \\ 0 & 0 & \frac{-2q^2}{K_{z,\mathrm{Vac}}^2(q,\omega)} \end{bmatrix}   e^{2iK_{z,\mathrm{Vac}}(q,\omega)h_1} \right\},
\end{align}
}where $K_{z,i}(q,\omega)=\sqrt{\epsilon_{\mathrm{r},i}(\omega)(\omega/c)^2-q^2}$ is the z-component wavevector in the media with the dielectric function $\epsilon_{\mathrm{r},i}(\omega)$ in Eq.~(\ref{Eq:dielectric}). $R_\mathrm{s}(q,\omega)$ and $R_\mathrm{p}(q,\omega)$ are the reflection coefficients of the s- and p-polarized electric fields, respectively, and they can be expressed as\cite{chew1995waves,Novotny2012}
\begin{align}
\label{Eq:Reflection_Coeff}
\nonumber
    &R_\mathrm{s(p)}(q,\omega) \\&= 
    \frac{r_\mathrm{s(p),\mathrm{Vac-NaCl}}(q,\omega)+r_\mathrm{s(p),\mathrm{NaCl-Ag}}(q,\omega)e^{2iK_{z,\mathrm{NaCl}}(q,\omega)h_2}}{1+r_\mathrm{s(p),\mathrm{Vac-NaCl}}(q,\omega)r_\mathrm{s(p),\mathrm{NaCl-Ag}}(q,\omega)e^{2iK_{z,\mathrm{NaCl}}(q,\omega)h_2}} ,
\end{align}
where $r_{\mathrm{s},i-j}(q,\omega)$ and $r_{\mathrm{p},i-j}(q,\omega)$ are the Fresnel reflection coefficients of the s- and p-polarized electric fields between $i$ and $j$ layer, respectively, and they can be expressed as\cite{chew1995waves,Novotny2012},
\begin{align}
    &r_{\mathrm{s},i-j}(q,\omega) = \frac{K_{z,i}(q,\omega)-K_{z,j}(q,\omega)}{K_{z,i}(q,\omega)+K_{z,j}(q,\omega)}, \label{Eq:s-Fresnel} \\
    &r_{\mathrm{p},i-j}(q,\omega) = \frac{\epsilon_{\mathrm{r},j}(\omega) K_{z,i}(q,\omega)-\epsilon_{\mathrm{r},i}(\omega) K_{z,j}(q,\omega)}{\epsilon_{\mathrm{r},j}(\omega) K_{z,i}(q,\omega)+\epsilon_{\mathrm{r},i}(\omega) K_{z,j}(q,\omega)} .
\label{Eq:Fresnel_coefficient_p}
\end{align}

Note that NaCl and silver are non-ferromagnetic substance, so that the permeabilities of different media in Eq.~(\ref{Eq:s-Fresnel}) are assumed to be equal.

\end{appendix}

\bigbreak

\end{document}